# TRANSFORMATION AND SIMULATION FOR A GENERALISED QUEUING PROBLEM USING A G/G/$n$/G/+ QUEUEING MODEL

BEN O'NEILL[*], *Deloitte Australia*[**]

WRITTEN 11 NOVEMBER 2021

**Abstract**

We examine a generalised queuing model which we call the G/G/$n$/G/+ model, which encompasses the G/G/$n$ and G/G/$n$/$s$ models as special cases. Our model accommodates useful generalisations in user behaviour and limitations on the facilities for the queuing process. Give a set of inputs for the users and facilities, we develop a recursive algorithm that computes all aspects of the queuing process, including the waiting-times, use-times and unserved-times for each user. We also show how the queue can be represented graphically in a "queuing plot". We use our algorithm to undertake simulation analysis to determine the distribution of queuing outputs given specified distributions for the inputs, and we show how this can be used to optimise the number of facilities. We conduct some simple simulations to illustrate the method using standard queuing models. Our method is implemented in various queuing functions in the `utilities` package in `R`.

QUEUING PROBLEMS; G/G/n/G/+ MODEL; EVALUATION METRICS; COMPUTATION; SIMULATION; R.

Queuing problems occur when users arrive intermittently at an amenity with a finite number of service facilities and users must wait for a service facility to become available due to its use by other users. There is a substantial literature on the statistical analysis of queuing problems including several major textbooks (see e.g., Stewart 2009; Khinchin 2013; Shortle et al 2018). The literature on queuing problems involves a variety of well-developed models and methods for analysing the waiting times of users and other metrics relating to the performance of the amenity (see e.g., Kingman 2009). Queuing models generally posit the joint probabilistic behaviour of a set of arrival-times and use-times for a set of users, and these are frequently framed in terms of continuous-time Markov chains. For models that are not too complicated, it is possible to obtain closed-form analytical results for the distributions of waiting-times and other important output metrics, but some models are too complicated for this. In the latter case, statisticians usually fall back on discrete-event-simulation (DES) models or other simulation methods to estimate the distributions of output metrics of interest (see e.g., Fishman 2001; Allen 2011; Choi and Kang 2013; Ghaleb, Suryahatmaja and Alharkan 2015).

In this paper we develop a recursive queuing algorithm to compute a range of outputs for users and service facilities in a highly general queuing model, using the arrival-times and (intended) use-times and other necessary variables as inputs. Our goal will be to convert these inputs into a specification of the queue outcome and resulting evaluation metrics. This will be used to allow simulation analysis of queuing metrics under any distributions that generate vectors of arrival-times, intended-use-times and other inputs. Our algorithm is less general than broader facilities for DES modelling, but it is far more parsimonious, and it is able to encompass a wide range of queuing models that are used in statistical practice.

---

[*] E-mail address: ben.oneill@hotmail.com.
[**] Data, Analytics and Technology Team, Deloitte, 8 Brindabella Circuit, Canberra ACT 2609, Australia

Our analysis will examine a highly generalised queuing process which (extending the general Kendall notation used in queuing models) we call a G/G/$n$/G/+ model.[1] The elements of this model corresponding to its notation are shown in Table 1 below. The main advantage of our model is that it encompasses both the G/G/$n$ queuing model and the G/G/$n$/$s$ queuing model as special cases, but it allows much broader user behaviour, where the maximum waiting-time for each user may be specified as a function of the queue-priority of that user. (More detail on this specification are given later in the paper.) This allows us to model limits on the size of the queue, but it also allows us to model more general user behaviour where users will leave the queue if they have not advanced to a desired queue-priority within a given waiting-time. The model also adds some additional generalisations that are useful in realistic queuing problems, including an allowance for revival-times for the facilities and closure-times for the amenity.

| | |
|---|---|
| **G**/G/$n$/G/+ | We allow an arbitrary input for the arrival-times of the users, which can be simulated from any distribution. |
| G/**G**/$n$/G/+ | We allow an arbitrary input for the intended use-times of the users, which can be simulated from any distribution. |
| G/G/***n***/G/+ | We allow any finite number of service facilities $n \in \mathbb{N}$. |
| G/G/$n$/**G**/+ | We allow an arbitrary input for the maximum waiting-time for each user, which will be a function that may depend on the queue-priority of the user. This generalises the G/G/$n$/$s$ model where users leave the amenity immediately if there are $s$ or more other users already in the queue when they arrive, and wait indefinitely otherwise. |
| G/G/$n$/G/**+** | We add additional generalisations including allowance for a fixed revival-time for the facilities (time needed for the facility to become available again after a user has finished using it) and allowance for closure-times on the amenity where it can close to new arrivals, close to new services, or terminate all existing services. |

**Table 1:** Elements of the G/G/$n$/G/+ model

---

[1] We use a slight abuse of notation in this name; the final element represented by the + sign signifies that we allow additional generalisations that are not easily accommodated within the Kendall notation. Thus, our model form can be read as a G/G/$n$/G model *plus other generalisations*.

The purpose of developing our algorithm is to allow simulation analysis where we can input an arbitrary set of inputs for the arrival-times, intended-use-times and maximum waiting-times (which we call patience-times). Since the algorithm can accommodate any values for these inputs this allows us to simulate an underlying queuing model with any distribution for these inputs. (Hence the Gs in our G/G/$n$/G/+ model.) This allows us to simulate the distribution of relevant output quantities in the model. Simulations can be used to give empirical confirmation to analytical results for G/G/$n$ and G/G/$n$/$s$ models, and it can also be used to approximate distributions and moment quantities for cases where an analytical solution is impracticable.

In addition to developing an algorithm to compute the outputs of the queuing process, we will show how the queue can be visualised using a "queuing plot" that shows the behaviour of each user and when each of the facilities is being used over the time-period under analysis. The queuing plot allows analysts to visualise the experience of each user and observe the relative sizes of the waiting-times, use-times and unserved-times for the set of users. To assist readers in applying our model and simulation method, our algorithm and resulting summary and plotting functions are all available in the **utilities** package in **R** (O'Neill 2021).

**1. Inputs and outputs for simulating the G/G/$n$/G/+ queuing model**

Consider an amenity containing $n$ service facilities and suppose $K$ users arrive at this amenity to use the service facilities according to some queuing model. We will let $\boldsymbol{t}_K = (t_1, t_2, \ldots, t_K)$ denote the **arrival-times** and $\boldsymbol{u}_K^* = (u_1^*, u_2^*, \ldots, u_K^*)$ denote the **intended-use-times** for these users. This means that user $k$ arrives at the amenity at time $t_k$ and intends to use the facility for $u_k^*$ time units (once a service facility becomes available for their use). Without any loss of generality, we assume that the vector of arrival-times is in non-decreasing order, so that the users $k = 1, \ldots, K$ are ordered from earliest arrival-time to latest arrival-time. Since the arrival-times, intended-use-times, and number of facilities are arbitrary inputs, this reflects the first three elements of the G/G/$n$/G/+ model.

To accommodate the user behaviour of our model, we stipulate that each user has a maximum-waiting-time (which we will call their "patience-time") that may depend on their priority in the queue. We let $\boldsymbol{w}_K^* = (w_1^*, w_2^*, \ldots, w_K^*)$ denote the **patience-times** of the users, where each of these patience-times is a non-increasing function $w_k^*: \mathbb{N} \to \mathbb{R}_+$ mapping the queue-priority of the user to the maximum amount of time they are willing to wait while at that priority. Each value $w_k^*(\kappa)$ represents the maximum waiting-time the user is willing to endure when he is at priority $\kappa$ in the queue — i.e., at each waiting time $0 < w_k^*(k) \leq \cdots \leq w_k^*(1)$ the user will leave the amenity unless their queue priority at that time is less than $\kappa$ (so the patience-time $w_k^*(1)$ is the maximum amount of time the user will wait). This form of user behaviour takes account of the queue priority, but it does not require the user to speculate on the use-times of other users or make any kind of probabilistic calculation in relation to their expected waiting-time. Since the patience-time is an input to the algorithm, this reflects the fourth element of the G/G/$n$/G/+ model. In particular, this input is sufficiently general to encompass simulation of the G/G/$n$/$s$ queuing model and is in fact far more general than that model.[2]

In order to determine when users are served, we stipulate some basic service rules. Only one user can use a service facility at a time, and users are allocated to facilities on a "first come, first served" basis. If multiple service facilities are available to a user, we assume that the first available facility is used, in the order $1, \ldots, n$.[3] Once a user has finished using a service facility there is a fixed **revival-time** $r \geq 0$ until that service facility is ready for a new user. We also allow the amenity to impose **closure-times** $0 < T_* \leq T_\circ \leq T_\bullet \leq \infty$ respectively for new arrivals, new service initiation, and termination of any existing services. (Note that one or more of these times can be set to infinity, which corresponds to there being no closure of the relevant kind at the amenity.) These inputs reflect the fifth element of the G/G/$n$/G/+ model.

Two important statistics we will examine are the amount of time each user waits for service, the amount of time each user is served, and the service facility that each user uses. To this end, we let $\boldsymbol{w}_K = (w_1, w_2, \ldots, w_K)$ denote **waiting-times** and we let $\boldsymbol{u}_K = (u_1, u_2, \ldots, u_K)$ denote the **use-times** for the users. There are several reasons that the actual use-time may not match the intended-use-time. Since users may potentially wait a long period of time for service, we will allow for the possibility that users might get bored of waiting and leave without service.

---

[2] In the G/G/$n$/$s$ model there is a maximum queue size of $s$. The users leave the amenity immediately if there are already $s$ other users in the queue when they arrive; otherwise they wait until they are served. This can be done within our input framework by setting the patience time to be:

$$w_k^*(\kappa) = \begin{cases} \infty & \text{if } \kappa \leq s, \\ 0 & \text{if } \kappa > s. \end{cases}$$

[3] Note that this allocation method favours the facilities with smaller numerical labels, so the outputs for the service facilities used are not expected to be exchangeable.

Another important statistic we will examine is the vector specifying which service facilities are used by each user. This output allows us to map the users to the service facilities, to see when the latter are in use. We let $\boldsymbol{F}_K = (F_1, F_2, \ldots, F_K)$ denote the **service-facilities** for the users; the service facility for each user is a number $1, \ldots, n$ for the service facility that gives service to the user (or NA if the user is not served at all). Combining this output with the use-time vector for the users allows us to obtain the use-times for the service facilities rather than the users. We let $\boldsymbol{U}_n = (U_1, U_2, \ldots, U_n)$ denote the **facility-use-time** for the service facilities, with elements (representing facility-use-time for a specific service facility) given by:

$$U_i = \sum_{k=1}^{K} u_k \cdot \mathbb{I}(F_k = i).$$

The outputs $\boldsymbol{w}_K$, $\boldsymbol{u}_K$ and $\boldsymbol{F}_K$ are service metrics for the users and the output $\boldsymbol{U}_n$ are service metrics for the service facilities. These are the primary outputs of interest. Combining these outputs with the initial inputs, for any user $k$ we have the time $t_k$ a user arrived at the amenity, the time $t_k + w_k$ at which service was initiated, and the time $t_k + w_k + u_k$ at which service ended and the user leaves the amenity (we refer to these as the **service-initiation-time** and **leaving-time**). We can also compute the start and end times of each service provided to a user at a particular service facility, though this is usually not particularly helpful.

In order to compute the statistics of interest pertaining to the queuing outcome, we will use some intermediate variables that are not of direct interest, but they are used to keep track of the state of the queuing system as time progresses. In particular, we will use a matrix of delay times for the services facilities at each arrival time for the users. We let $\mathbf{D} = [D_{k,i}]$ denote the **delay matrix** for the system, indexed over $k = 0, \ldots, K$ and $i = 1, \ldots, n$. The value $D_{k,i}$ is the delay until service becomes available at service facility $i$, measured from time $t_k$.[4] The delay matrix is a key object to track to compute the waiting-times for users when they arrive at the amenity. At each arrival-time the new user is allocated to the service facility with the lowest delay (using a lower service facility number in the case of a tie). When the user is allocated to the service facility the delay matrix is updated for the next time period, with the time taken by that user —and the revival time— being added to the delay for the allocated service facility.

---

[4] The delay $D_{k,i}$ is measured *including* the use-time of the user who arrives at time $t_k$. For recursive reasons, the delay matrix includes the index $k = 0$, which refers to the starting time $t_0 = 0$. We assume that all service facilities are available at the starting time so we take $D_{0,i} = 0$ for all $i = 1, \ldots, n$.

We will also need some intermediate variables to determine which users leave the queue prior to being served, due to the wait exceeding the relevant patience-time. To do this, consider the waiting-time problem for user $k$ who arrives at time $t_k$. When this user arrives at the amenity, all $k-1$ previously arriving users will be ahead of him in the queue unless they have already exited the queue (either because they have begun service or because they have left without service). For each previous user $1 \leq i < k$, the time at which that user leaves the queue is $t_i + w_i$. Consequently, at time $t_k + w$ the present user has waited $w$ time-units since arrival and each previous user $i$ is ahead of him in the queue if and only if $t_i + w_i > t_k + w$. This allows us to define the **queue-priority function** for user $k$ by:

$$\Pi_k(w) \equiv 1 + \sum_{i=1}^{k-1} \mathbb{I}(t_i + w_i > t_k + w) \qquad \text{for all } w \geq 0.$$

The value $\Pi_k(w)$ is the **queue-priority** for user $k$ after he has waited $w$ time units — it is equal to one plus the number of previous users who are still in the queue at that time (and who are therefore ahead of the present user in the queue). Since our algorithm is recursive, we have already computed all relevant queuing metrics for the previous users, so the function $\Pi_k$ will be a known quantity when we do the computations for user $k$.

Now, given this function, we can figure out the maximum waiting-time for the user given any patience-time function $w_k^*$. Suppose that the user is still in the queue up to the waiting time $w_k^*(\kappa)$ for some $\kappa = 1, \ldots, k$. The user will leave the queue at that time if $\Pi_k(w_k^*(\kappa)) \geq \kappa$ (i.e., if they have reached their patience-time for queue-priority $\kappa$). Setting aside leaving due to the closure of the facilities, the user will leave the queue the first time this occurs. This means that their **maximum waiting-time** is $w_k^*(\hat{\kappa})$ where the **maximum queue-priority** $\hat{\kappa}_k$ is given by:

$$\hat{\kappa}_k \equiv \max\{\kappa = 1, \ldots, k | \Pi_k(w_k^*(\kappa)) \geq \kappa\}.$$

The value $w_k^*(\hat{\kappa})$ represents the maximum waiting-time for the user, setting aside leaving due to closure of the facilities, and so it effectively replaces the single patience-time that was used in the previous recursive method.

To simplify the analysis, we can observe that the queuing process depends on each patience-time function $w_k^*$ only at the points $w_k^*(1), \ldots, w_k^*(k)$. This means that all relevant information for the patience-times for a queuing process with $K$ arrivals is encapsulated in a lower-triangle matrix $\boldsymbol{w}^* = [w_k^*(\kappa)]$ with indices taken over the range $k = 1, \ldots, K$ and $\kappa = 1, \ldots, K$. In our recursive algorithm we will use this input along with the arrival-times and waiting-times for the previous users to form the corresponding lower-triangle matrix $\mathbf{\Pi}^* = [\Pi^*_{k,\kappa}]$ with elements $\Pi^*_{k,\kappa} \equiv \Pi_k(w^*_k(\kappa))$ giving the **forward-queue-priority** of user $k$ at the expiry of their patience time for queue-priority $\kappa$. Once we have recursively computed the values $\Pi^*_{k,1}, \dots, \Pi^*_{k,k}$ for user $k$ we then obtain the simplified formula $\hat{\kappa} = \max\{\kappa = 1, \dots, k | \Pi^*_{k,\kappa} \geq \kappa\}$.

In Table 2 below we show the variables in the queuing problem. Inputs describing the user behaviour are shown in yellow and inputs describing the service-facilities and their behaviour are shown in orange. Intermediate variables are shown in red and the outputs determined by interaction of users and service facilities (which we want to compute) are shown in blue.

| Description | Variable |
|---|---|
| Arrival-time* | $t_k$ |
| Intended-use-time* | $u^*_k$ |
| Patience-time (maximum waiting-time)* | $w^*_k$ |
| Number of facilities | $n$ |
| Revival-time | $r$ |
| Closure-time for new arrivals | $T_*$ |
| Closure-time for new services | $T_\circ$ |
| Closure-time for terminating existing service | $T_\bullet$ |
| Service indicator* | $\delta_k$ |
| Delay until service facility is available* ** | $D_{k,i}$ |
| Forward-queue-priorities* *** | $\Pi^*_{k,\kappa}$ |
| Maximum queue-priority* | $\hat{\kappa}_k$ |
| Waiting-time* | $w_k$ |
| Use-time* | $u_k$ |
| Service-initiation-time* | $t_k + w_k$ |
| Leaving-time* | $t_k + w_k + u_k$ |
| Unserved-time* | $u^*_k - u_k$ |
| Service-facility used by the user* | $F_k$ |
| Use-time for the service facility** | $U_i$ |
| \* User variables are taken over the indices $k = 1, \dots, K$ | |
| \*\* Service facility variables are taken over the indices $i = 1, \dots, n$ | |
| \*\*\* The queue-priority is taken over the indices $\kappa = 1, \dots, k$ | |

**Table 2:** Variables used in simulation of the G/G/$n$/G/+ model

## 2. A recursive method for computing the queuing output and evaluation metrics

To compute the outputs of the queuing process, we recursively update the user information in the order of their arrival and keep track of the state of the system with intermediate variables. For each user $k = 1, \ldots, K$ we check to see if the user has arrived before the closure-time for new arrivals. If $t_k \geq T_*$ (i.e., the amenity is closed to new arrivals) then we update the queue metrics as follows. (We take $t_0 \equiv 0$, $D_{0,i} \equiv 0$, etc. for indices outside the allowable range.)

| | | |
|---|---|---|
| Set waiting time | $w_k = 0$ | |
| Set use time | $u_k = 0$ | |
| Set service indicator | $\delta_k = 0$ | |
| Set service facility | $F_k = \text{NA}$ | |
| Set delay vector | $D_{k,i} = \max(D_{k-1,i} - (t_k - t_{k-1}), 0)$ | |

Contrarily, if $t_k < T_*$ (i.e., the amenity is open to new arrivals) then we update as follows:

| | |
|---|---|
| Set provisional delay times | $\widehat{D}_{k,i} = \max(D_{k-1,i} - (t_k - t_{k-1}), 0)$ |
| Set provisional waiting time | $\widehat{w}_k = \min(\widehat{D}_{k,1}, \ldots, \widehat{D}_{k,n})$ |
| Find queue-priorities | $\Pi^*_{k,\kappa} = \Pi_k(w^*_k(\kappa))$ |
| Set maximum queue-priority | $\hat{\kappa} = \max\{\kappa = 1, \ldots, k \mid \Pi_k(w^*_k(\kappa)) \geq \kappa\}$ |
| Set maximum waiting time | $w^{**}_k = \min(w^*_k(\hat{\kappa}), T_\circ - t_k)$ |
| Set waiting time | $w_k = \min(\widehat{w}_k, w^{**}_k)$ |
| Set service indicator | $\delta_k = \mathbb{I}(\widehat{w}_k < w^{**}_k)$ |
| Set use time | $u_k = \delta_k[\min(w_k + u^*_k, T_\bullet - t_k) - \min(w_k, T_\bullet - t_k)]$ |
| Set service facility | $F_k = \begin{cases} \text{NA} & \text{if } \delta_k = 0 \\ \operatorname{argmin}(\widehat{D}_{k,1}, \ldots, \widehat{D}_{k,n}) & \text{if } \delta_k = 1 \end{cases}$ |
| Set delay vector | $D_{k,i} = \widehat{D}_{k,i} + \delta_k(u_k + r)\mathbb{I}(i = F_k)$ |

**ALGORITHM 1: Simulation of the G/G/n/G/+ queuing process**

**Inputs:**
**t**         Vector of arrival-times (non-decreasing non-negative values)
**u***        Vector of intended-use-times (vector of non-negative values)
**w***        Matrix of patience-times (matrix of non-negative values)
**n**         Number of service facilities (positive integer)
**T***        Closure-time for new arrivals (positive numeric value or Inf)
**T₀**        Closure-time for new services (positive numeric value or Inf)
**T.**        Closure-time for all services (positive numeric value or Inf)

**Outputs:**
**w**         Vector of waiting-times
**u**         Vector of use-times
**F**         Vector of service facilities
**U**         Vector of use-times for service facilities

```
#Set output vectors
K <- length(t)
w <- [0, …, 0]    (length K)
u <- [0, …, 0]    (length K)
F <- [NA, …, NA]  (length K)
U <- [0, …, 0]    (length n)

#Set delay matrix and forward-queue-priority matrix
D  <- Matrix with rows indexed by k = 0,…,K and columns indexed by i = 1,…,n
QP <- Matrix with rows indexed by k = 1,…,K and columns indexed by κ = 1,…,K
      (Initial values in both matrices are all set to zero)

#Compute user outputs recursively
TIME <- 0
for each k = 1,…,K {

  #Update delays/time
  for each i = 1,…,n { D[k, i] <- max(D[k-1,i] – (t[k]-TIME), 0) }
  TIME <- t[k]

  #Update user statistics
  if (t[k] ≥ T*) {
    w[k]  <- 0
  } else {
    ŵ     <- min(D[k, ])
    end.t <- t[1:(k-1)] + w[1:(k-1)]
    for each κ = 1,…,k { QP[k, κ] <- 1 + sum(end.t > t[k] + w*[k,κ]) }
    k.hat <- Maximum value of κ = 1,…K with QP[k, κ] ≥ κ
    w.max <- w*[k, k.hat]
    w**   <- min(w.max, T₀-TIME)
    w[k]  <- min(ŵ, w**)
    if (ŵ < w**) {
       F[k]    <- smallest index of D[k, ] giving its minimum value }
       u[k]    <- min(w[k]+u*[k], T.-TIME) - min(w[k], T.-TIME)
       D[F[k]] <- D[F[k]] + u[k] + r } }
#Compute service facility outputs
for each i = 1,…,n {
  IND  <- indicator(F = i)   (vector of indicator values)
  U[i] <- sum(u*IND) }

#Return outputs
list(w, u, F, U)
```

Our recursive method is implemented in Algorithm 1 above; the algorithm can be "vectorised" if needed, to allow inputs from multiple simulations. The algorithm gives outputs for the waiting-times and use-times for the users, the service facilities for each user, and the use-times for each service facility. These outputs can be used to compute the service-initiation-time, leaving-time and unserved-time for each user, which then gives all the outputs shown in blue in Table 2. While the goal here is to calculate the output variables shown in blue, it is possible to combine these variables with the initial input variables to create tables showing all relevant times for users and service facilities in the queuing problem. It is also possible to represent this information graphically to aid user interpretation and give an idea of the dynamics of the system. This kind of presentation gives a holistic set of information for the queuing process, which shows how convenience it is to users and how well the service facilities are used.

The present algorithm has been implemented to give full output of all information for the users and facilities in the `queue` function in the `utilities` package in `R` (O'Neill 2021). This function allows the user to specify each input highlighted in yellow and orange in Table 2 (with default values for some inputs if they are not specified). It produces a table of user information and a table of service facility information. This object can also be plotted with the `plot` function to produce a "queuing plot". The description and inputs for the `queue` function and some related functions in the package are shown in Table 3 below.

| Function | Description | Inputs |
|---|---|---|
| `queue` | Produces a `queue` object containing information on the users and service facilities in the queuing process. | `n, arrive, use.full,`<br>`wait.max = NULL, revive = 0,`<br>`close.arrive = Inf,`<br>`close.service = Inf,`<br>`close.full = Inf` |
| `plot.queue` | Generates a queuing plot from a `queue` object. This plot shows the user metrics and the service facility metrics. | `OBJECT*, print = TRUE,`<br>`gap = NULL,`<br>`line.width = 2,`<br>`line.colours = NULL, ...` |
| `summary` | Generates a summary of a `queue` object containing various summary statistics for the process. | `OBJECT*`<br>`probs = NULL,`<br>`probs.decimal.places = 2` |
| `plot.summary.queue` | Generates a queuing summary plot from a `summary.queue` object. This plot shows histograms of user metrics. | `OBJECT**, print = TRUE,`<br>`count = FALSE,`<br>`bar.colours = NULL, ...` |
| * The input for this function must be a `queue` object ||| 
| ** The input for this function must be a `summary.queue` object ||| 

Table 3: Functions relating to the `queue` function in the `utilities` package.

It is simple to use the **queue** function to generate comprehensive queuing information from the required inputs. In the code below we simulate a queuing problem with $K = 20$ users and $n = 3$ service facilities (full code in supplementary materials). We use random arrival-times and intended-use-times using a Markov model with constant rates (i.e., we use the exponential distribution) and we set the patience-times and revival-time to fixed values. We then use the **queue** function to generate and print the queuing information, and we give the queuing plot for the process in Figure 1. (Our abridged code is in blue and the output is in black.)

```
#Compute and print queuing information with n = 3
QUEUE <-queue(arrive = ARRIVE, use.full = USE.FULL,
              wait.max = WAIT.MAX, n = 3, revive = 2,
              close.arrival = 30, close.full = 35)
QUEUE

#Display the queuing plot
plot(QUEUE)

      Queue Information

Model of an amenity with 3 service facilities with revival-time 2
Service facilities close to new arrivals at closure-time = 30
Service facilities close to new services at closure-time = 35
Service facilities end existing services at closure-time = 35

Users are allocated to facilities on a 'first-come first-served' basis

------------------------------------------------------------------------

User information

            arrive      wait       use     leave   unserved  F
user[1]   1.132773 0.000000 1.295324   2.428096 0.0000000  1
user[2]   2.905237 0.000000 8.684531  11.589768 0.0000000  2
user[3]   3.123797 0.000000 4.935293   8.059090 0.0000000  3
user[4]   3.333490 1.094606 9.851356  14.279452 0.0000000  1
user[5]   3.987593 3.032653 0.000000   3.987593 7.7647223 NA
user[6]   8.330046 1.729045 9.395193  19.454283 0.0000000  3
user[7]  10.174389 3.032653 0.000000  10.174389 6.6364357 NA
user[8]  10.983913 1.839397 0.000000  10.983913 6.3566350 NA
user[9]  12.418764 1.171004 9.472275  23.062043 0.0000000  2
user[10] 12.639333 3.032653 0.000000  12.639333 0.2799744 NA
user[11] 14.725436 1.554016 5.726761  22.006213 0.0000000  1
user[12] 15.868481 3.032653 0.000000  15.868481 8.7877649 NA
user[13] 17.724886 3.032653 0.000000  17.724886 8.3127787 NA
user[14] 24.360787 0.000000 5.731435  30.092223 0.0000000  1
user[15] 25.942602 0.000000 9.057398  35.000000 1.2771158  2
user[16] 27.495468 0.000000 5.257165  32.752633 0.0000000  3
user[17] 30.309521 0.000000 0.000000  30.309521 2.9375673 NA
user[18] 31.291641 0.000000 0.000000  31.291641 0.8481486 NA
user[19] 31.797041 0.000000 0.000000  31.797041 1.1935939 NA
user[20] 32.679761 0.000000 0.000000  32.679761 3.7952605 NA

------------------------------------------------------------------------
```

```
Facility information

      open end.service      use revive
F[1]     0    30.09222 22.60488      8
F[2]     0    35.00000 27.21420      6
F[3]     0    32.75263 19.58765      6
```

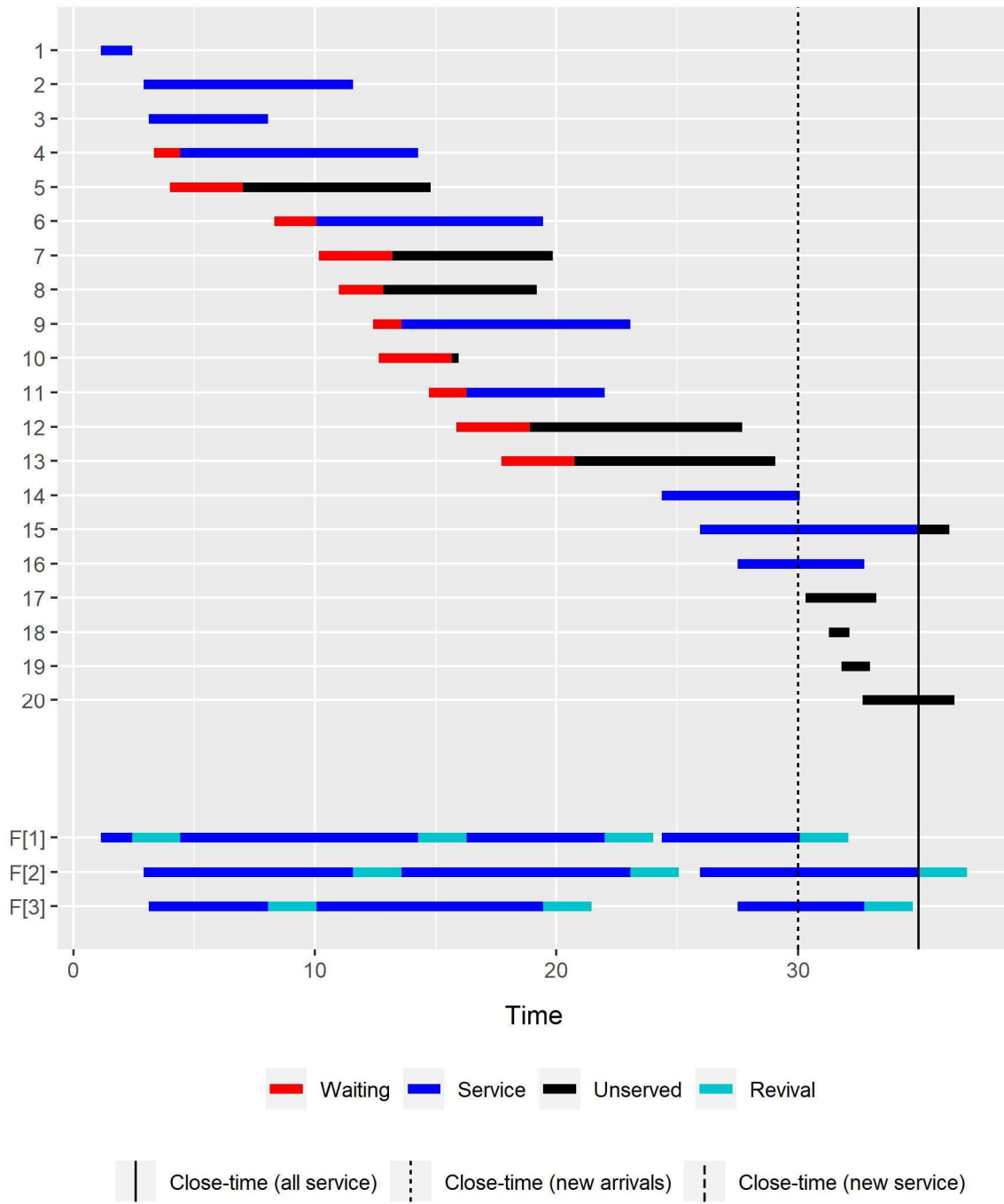

**Figure 1:** Queuing plot for the process shown in code above;
produced using the `queue` function in the `utilities` package.

The queuing plot in Figure 1 allows us to visualise the effectiveness of the amenity in servicing the users. In particular, the waiting-times and the unserved-times both constitute negatives for the user. A similar type of plot was examined in Ingolfsson and Grossman Jr (2002), but our version is a bit more general. We can see from the plot that users $k = 5, 7, 8, 10, 12, 13$ wait up to their patience-time but leave without being served. We also see that users $k = 17, \ldots, 20$ arrive after the closure-time for new arrivals and so they leave immediately without being served. The former is a sign that the number of service facilities may be inadequate for the volume of users in the process. In the bottom part of the plot we see the total use-time and revival-time for the service facilities. There is some idleness represented by gaps between the lines. This shows that the service facilities have some spare capacity at some times, but they still do not succeed in serving all the users who arrive before the closure time.

After having created a queuing object using the `queue` function we can produce summary information for the users, showing the mean and standard deviation of the wait-times, use-times, unserved-times, intended-use-times and use-proportions, as well as quantiles of these values at specified probabilities. This is done using the `summary` function. In the code below we produce summary statistics for the queuing process and the queuing summary plot, which is shown in Figure 2. (Our code is in blue and the output is in black.)

```
#Display the summary statistics for the queue
SUMMARY <- summary(QUEUE)
SUMMARY

#Display the queueing summary plot
plot(SUMMARY)

    Summary Statistics for a Queuing Process

Model of an amenity with 3 service facilities with revival-time 2
Service facilities close to new arrivals at closure-time = 30
Service facilities close to new services at closure-time = 35
Service facilities end existing services at closure-time = 35

Users are allocated to facilities on a 'first-come first-served' basis

-----------------------------------------------------------------------

                     wait        use  unserved    use.full   use.prop
    Mean        1.1275667  3.4703365 2.4094999   5.8798364  0.4938211
    Std.Dev     1.2960294  4.0454886 3.2593190   3.2562273  0.5073632
    Quantile[0.00] 0.0000000 0.0000000 0.0000000 0.2799744  0.0000000
    Quantile[0.25] 0.0000000 0.0000000 0.0000000 3.5808372  0.0000000
    Quantile[0.50] 0.5473032 0.6476618 0.5640615 6.0440352  0.4382111
    Quantile[0.75] 2.1377112 6.4697094 4.4356041 8.7103397  1.0000000
    Quantile[1.00] 3.0326533 9.8513555 8.7877649 10.3345137 1.0000000
```

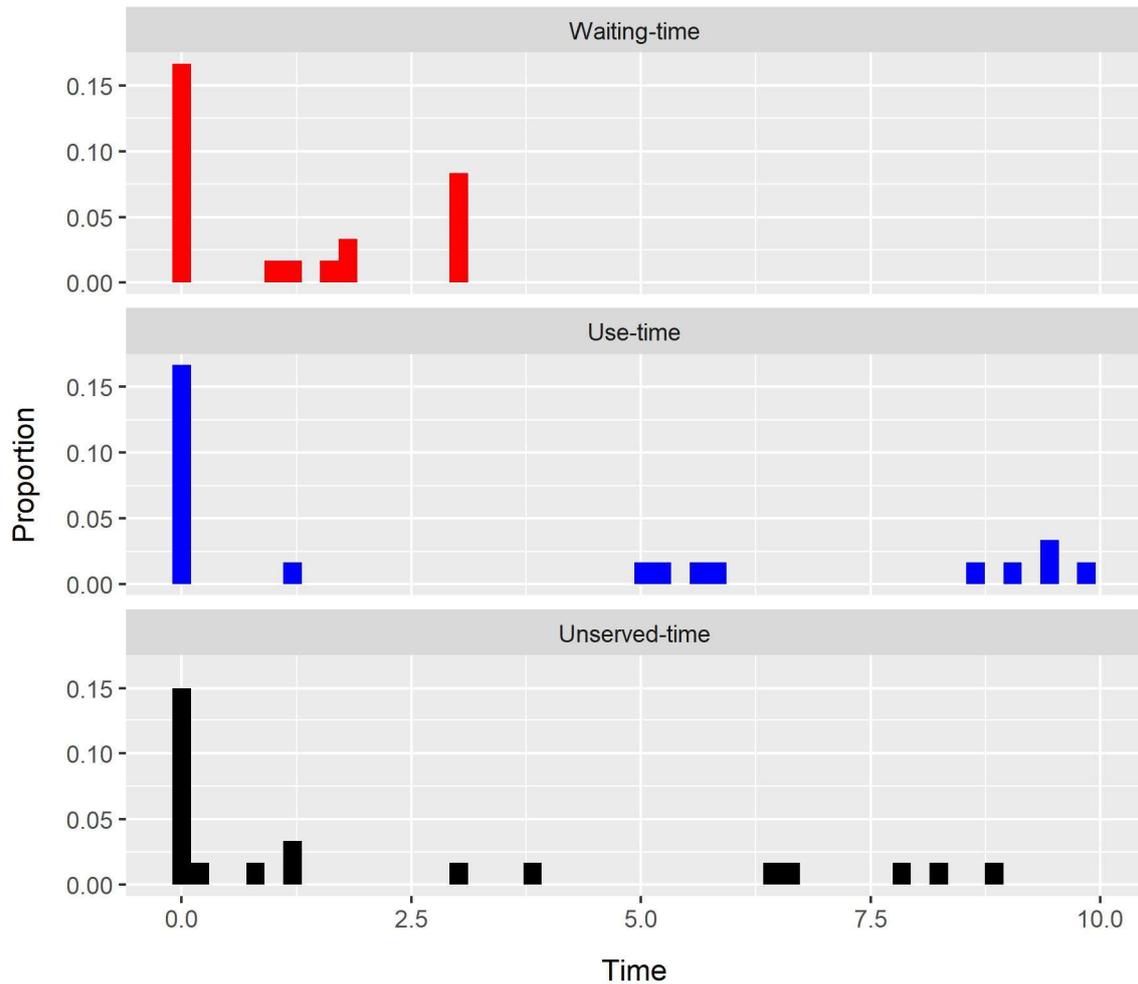

**Figure 2:** Queuing summary plot for the process shown in code above; produced using the `summary.queue` function in the `utilities` package.

The queuing summary plot in Figure 2 shows that there is some waiting by the users. We can see that the use-time is roughly commensurate with the unserved-time. Since there are only sixteen users who arrive before the closure-time for new arrivals, we have only a relatively small amount of data in this case. The information in Figures 1-2 can be combined to give an overview of the queuing process and give us an understanding of how well the users were served by the service facilities. If we were to use the same user inputs but change the number of service facilities this would lead to a change in the ability of the amenity to serve these users, with a consequent change in the outputs of the queuing process. In this case the large amount of waiting and unserved-time suggests that we may need to increase the number of facilities in the amenity.

## 3. Simulation and optimisation using the queue information

The above algorithm and resulting queuing functions allow us to undertake simulation analysis of queuing problems. In particular, we can undertake a form of analysis where the statistical model for the queuing process is used only to produce the simulations for the inputs, but all remaining analysis is done using simple model-free statistical methods. An obvious application is for determining the number of service facilities required to adequately service a set of users with queuing inputs determined by a particular generative model.

In the present section we will illustrate simulation and optimisation of the queuing process from a **generative model** for the inputs to the queuing process. For our generative model we will use Markovian arrival times with constant rate $\lambda$ and uniform intended-use-times with constant mean $\mu$. We will also assume that the maximum-waiting-times decay exponentially with the queue-priority $\kappa$. This gives the model form:

$$t_k = \sum_{\ell=1}^{k} \Delta_\ell \qquad \Delta_\ell \sim \text{IID Exp}(\lambda) \qquad u_k^* \sim \text{IID U}(0, 2\mu) \qquad w_k^*(\kappa) = \alpha \exp\left(-\frac{\kappa}{\beta}\right).$$

Our generative model will also use the revival-times and closure-times:

$$r = 2 \qquad T_* = 30 \qquad T_\circ = 35 \qquad T_\bullet = 35.$$

For our simulations we will use the following parameters for our analysis:

$$\lambda = 1.5 \qquad \mu = 6 \qquad \alpha = 5 \qquad \beta = 2.$$

While we use the present generative model for illustrative purposes, it is important to stress that the simulation method allows us to use any distributions of inputs for our generative model, giving a form of analysis that is applicable to any G/G/$n$/G/+ queuing model.

Using the algorithm formulated in this paper —and implemented in the `queue` function— we simulate the queue process using the stipulated generative model (full code in supplementary materials) with $M = 10^4$ simulations. Figure 3 below shows the simulated distributions of the mean-wait-time, mean-unserved-time and mean-use-time, each taken over all the users in the queue process in each simulation. This gives us a simple graphical understanding of the effect of varying the number of facilities in the amenity. As can be seen from the plot, if we increase the number of facilities the mean wait-time and mean unserved-time both decrease and the mean use-time increases, all of which accords with intuition. It is simple to compute the means of the distributions to see the change in means as we vary the number of service facilities.

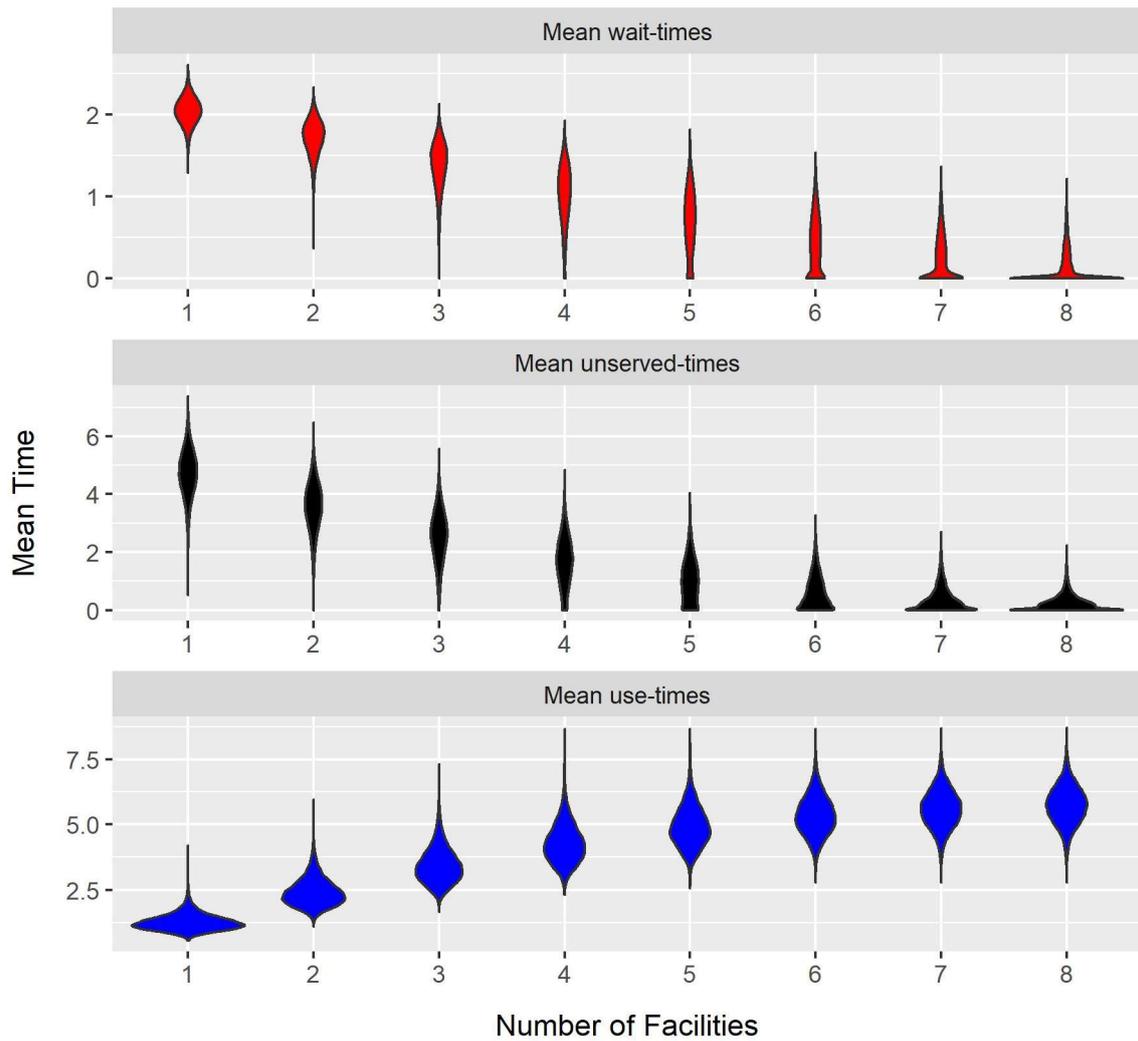

**Figure 3:** Plot of distributions of mean wait-times, unserved-times and use-times
This plot used $M = 10^4$ simulations of the queueing model.

Graphical inspection of the distributions of mean times allows a queuing analyst to determine a reasonable value for the number of service facilities. This can be done on an intuitive basis by looking at the change in the distributions of the mean-wait-times and mean-unserved-times. However, if we wish to reduce this to an unambiguous optimisation problem, we must stipulate cost information and minimise the resulting risk. A reasonable candidate for a loss function for this purpose is to have fixed costs for the facilities, total-wait-time and total-unserved-time (only for users arriving before the closure for new arrivals). Given the **facility-cost** $C_F > 0$, the **wait-cost** $C_w > 0$ and the **unserved-cost** $C_u > 0$ this **loss function** has the following form:

$$\text{Loss}(n, \boldsymbol{t}, \boldsymbol{u}^*, \boldsymbol{w}^*) = C_F n + \sum_{k=1}^{K} C_w w_k + \sum_{k=1}^{K} C_u (u_k^* - u_k) \mathbb{I}(t_k \leq T_*).$$

The first component of this loss function is the cost of the facilities in the amenity, the second component is the cost (in terms of inconvenience) for users waiting for services, and the third component is the cost (in terms of inconvenience) for users having unfulfilled services even after visiting the amenity. For this last term we adopt the rule that there is no "loss" for a user who arrives after the closure-time for new arrivals for the amenity.[5] Note that the inputs to the loss function are the value $n$ and the user inputs to the queuing problem, but the loss itself depends on the resulting outputs (which we can compute using our recursive method).

Optimisation of the appropriate number of service facilities involves minimisation of the risk (expected loss) under a specified generative model for the queuing inputs. For this purpose, suppose we generate $M$ simulations for the vectors $\boldsymbol{t}^{(\ell)}$, $\boldsymbol{u}^{*(\ell)}$ and $\boldsymbol{w}^{*(\ell)}$ (each with lengths $K^{(\ell)}$) over the indices $\ell = 1, \ldots, M$. These vectors can be simulated using any kind of generative model, so our analysis is agnostic to the particular distributions at issue. We can obtain a reasonable estimate of the risk (expected loss) by taking a large value of $M$ and computing:

$$\widehat{\text{Risk}}(n, \boldsymbol{t}, \boldsymbol{u}^*, \boldsymbol{w}^*) = \frac{1}{M} \sum_{\ell=1}^{M} \left[ C_F n + \sum_{k=1}^{K} C_w w_k^{(\ell)} + \sum_{k=1}^{K} C_u (u_k^{*(\ell)} - u_k^{(\ell)}) \mathbb{I}(t_k^{(\ell)} \leq T_*) \right].$$

By computing this estimated risk for values $n = 1, \ldots, N$ (up to some maximum value $N$ that is above the true optima[6]) we can get an accurate estimate of the true optimising value for $n$. In Figure 4 below we show the simulated distributions of the loss using our previous generative model with the stipulated costs $C_F = 30$, $C_w = 1$ and $C_u = 2$. This imposes a substantial fixed cost for each facility, with a small cost for the wait-times and a larger cost for the unserved-times. (Note that the loss function is determined by the *total* waiting-time and unserved-time, not the mean waiting-time and unserved-time. Consequently, these values are affected by the number of users at the amenity in each simulation.) In this example, the risk is minimised by using $n = 5$ service facilities for the amenity. Note that this computation is highly sensitive to the stipulated costs — in particular, if we increase the fixed cost of an additional facility then the optimal number of facilities will be reduced and if we reduce the fixed cost of an additional facility then the optimal number of facilities will increase.

---

[5] Although there is a user inconvenience in this case, in the present analysis we take the view that it is the user's own fault (due to failure to adhere to the advertised closure-time for the amenity), so it should not count as part of the loss. It is possible to conduct an alternative form of analysis where the closure-times are also variables that can be optimised, and in this latter case it is appropriate to include losses of this kind in the loss function.

[6] Once a value of $n$ is sufficient to serve all the users without waiting in the vast majority of simulations, taking a higher value will tend to increase the risk by adding to the cost of the facilities without any commensurate reduction in the costs for user waiting-times and unserved-times. In practice it is simple to see when this occurs.

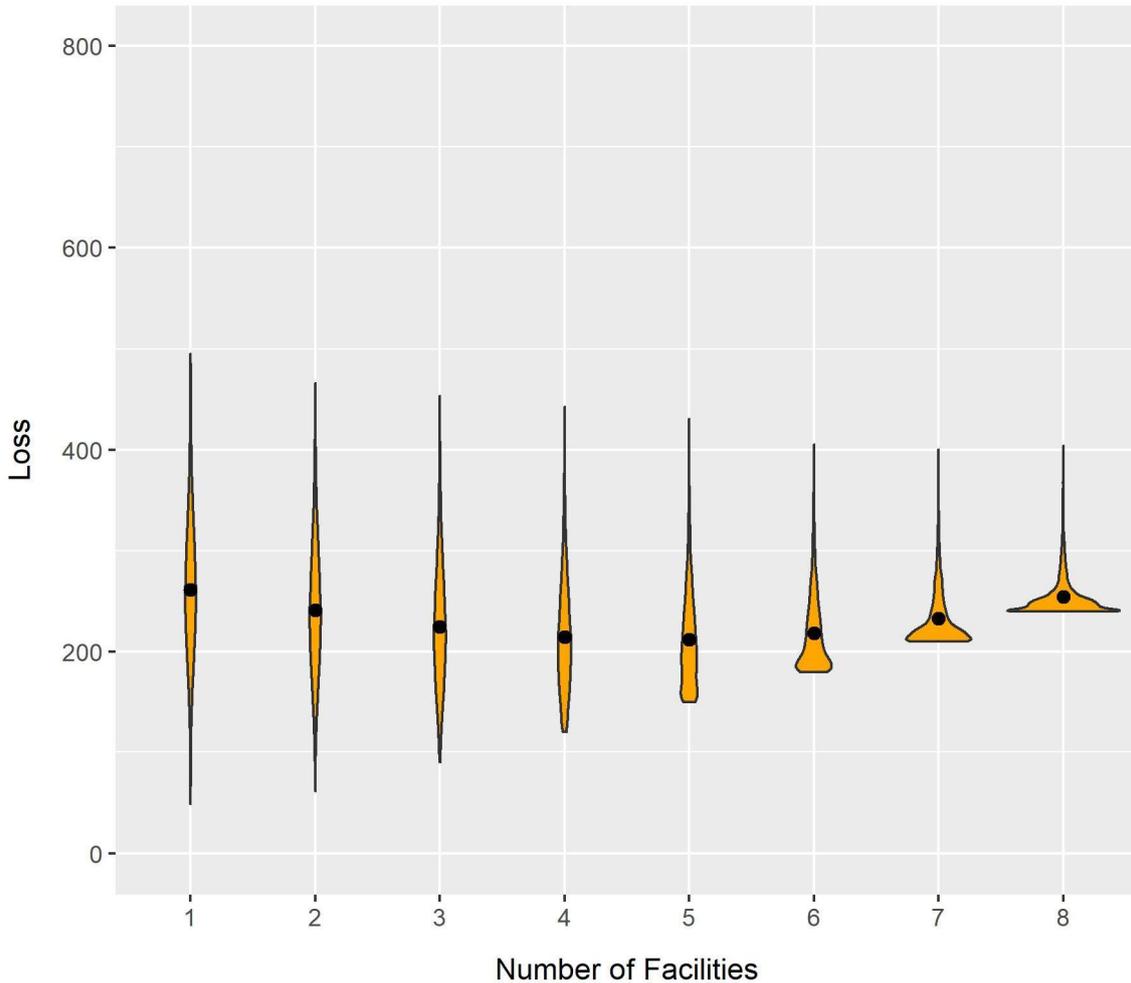

**Figure 4:** Plot of the loss distributions by number of facilities
This plot used $M = 10^4$ simulations of the queueing model.

Our simulation example here is used to demonstrate how we can undertake queueing analysis and make decisions about the appropriate number of service facilities under the highly general form of the G/G/$n$/G/+ queuing model. As previously noted, this general model encapsulates the G/G/$n$ and G/G/$n$/$s$ queuing models as special cases. Simulation is particularly useful in these cases because the models are sufficiently complicated that analytical results for the mean-times and total-times for important outputs are not available. (There are some approximate results in the literature for such models, but they are unable to accommodate generalisations in user behaviour like priority-dependent maximum-waiting-times. For example, approximate solutions for the moments of the outputs of G/G/$n$ models are in Halachimi and Franta (1977) and Pourbabai (1989).)

Our present simulation analysis involved an example with finite closure-times for the facilities, leading to a finite time-period where service is available. In some simulations the analyst is instead concerned with finding the "steady state" behaviour of the queue under the assumption that it runs forever. This latter problem can also be approximately solved using our method, by running simulations over a time-period that is sufficient for close convergence to the steady state of the system. For such purposes we recommend that the analyst simulate the queuing process over a substantial time period (with all closure-times set to be infinite) and have a "burn-in" period at the start of the queueing process that is eliminated when determining the simulated distributions of the outputs. Elimination of the "burn-in" period is used to avoid bias in the estimation problem, owing to the fact that all service facilities are available at the start-time for the queuing process.

**4. Possible extensions for more complicated user behaviour and facility specifications**

The queuing algorithm we have put forward in this paper is suitable for any underlying model within the $G/G/n/G/+$ model form. This is already quite a general model, and it encompasses a number of important queuing models in the statistical literature. Notwithstanding this level of generality, it is worth examining some further ways in which the model and algorithm could be extended, to deal with possible generalisations that reflect realistic variants in user behaviour or facility specifications. In this section we examine some possible extensions.

The $G/G/n/G/+$ model form is agnostic to the input distributions for the arrival-times, intended-use-times and maximum-waiting-time functions. Our algorithm for computing the outputs of this model allows easy simulation analysis with any distributions and functional forms for these input quantities. In particular, we can accommodate cases where there are periods of heavier and lighter intensity of user arrivals. As a general rule, if a model with constant arrival-intensity is replaced with one that has variable arrival-intensity (i.e., periods of light and heavy intensity) then *ceteris paribus* the periods of high intensity of arrivals this will lead to longer wait-times and unserved-times. This is because arrivals will tend to "bunch up" in time periods with heavy arrival-intensity, leading to overloading of the service facilities and consequent waiting (and possible abandonment) by the users. Due to variations in the arrival intensity in a queuing process, and the possibility of periods with higher arrival intensity, the provider of an amenity may wish to vary the number of service facilities available over different time periods. Our model and algorithm assumes that all service facilities operate over the same time period — i.e., they all have the same opening-times and closure-times— but it is possible to extend our algorithm to allow different facilities to have different opening-times and closure-times. For example, we may wish to allow the amenity to add service facilities during times with higher arrival-intensity and remove service facilities during periods with lighter arrival-intensity.

Extending the present algorithm to allow different opening-times for the facilities is trivial. The opening-times for the service facilities in our algorithm are given by the starting delay times $\mathbf{D}_0 = [D_{0,1}, \ldots, D_{0,n}]$ (i.e., the first row of the delay matrix). In our base algorithm these values are all set to zero, which means that all service facilities open at time $t = 0$. However, we can easily specify later opening-times (including different opening times for different service facilities) by taking $\mathbf{D}_0$ as an additional input in the algorithm specifying the first row of the delay matrix. The initial delay $D_{0,i} \geq 0$ then gives the specified opening-time for service facility $i$. (Note that we should set $\min D_{0,1}, \ldots, D_{0,n} = 0$ in this input so that the starting time $t = 0$ represents the time at which the first service facility opens.[7]) It is likewise possible to extend the analysis to allow for different closure-times. The extension essentially requires us to form closure-times $0 < T_{i,*} \leq T_{i,\circ} \leq T_{i,\bullet} \leq \infty$ for each service facility $i = 1, \ldots, n$. This generalisation leads to some minor changes in the recursive algorithm, whereby we compute the maximum waiting time and the service facility used by each user by taking account of the individual closure-times of each facility. It is possible (but complicated) to extend this further to allow facilities to reopen and close multiple times. Such an extension would require a broad set of inputs for the opening-times and closure-times of the facilities and would lead to some quite complicated adjustments to our recursive algorithm.

Another possible extension that may be useful in some cases is to specify more complicated behaviour for the revival-times for the service facilities. Our model and algorithm use a fixed revival time for all facilities, but this can be varied to allow different revival-times for different facilities, or stochastic rather than deterministic revival-times. It is also possible to extend the analysis to make the revival-times depend upon the prior use-time for the service facility — e.g., if the facility is used for longer by a user then the subsequent revival-time for that facility is longer. There are innumerable ways in which this aspect of the analysis could be extended, most of which would require simple variations to our algorithm. Such an extension would replace our fixed revival-time $r$ with times that are specified either by a vector of fixed values for each facility, or a new step in the algorithm determining each revival-time using a stochastic procedure.

---

[7] Note that if it is possible for users to arrive prior to the opening of the first service facility then the model should specify the behaviour of the user in this case. By default, the user will wait until their patience-time and then leave if they are not yet being served. One could vary the algorithm to have the user leave earlier, and this might be realistic in some cases. For example, if the opening times of the service facilities are known to the users (e.g., they are advertised at the amenity) then the user might leave immediately if the delay for the first open facility is known to be above their patience time.

In considering extensions to our model and algorithm, researchers should be cognisant of the benefits of having a deterministic function to map the queuing inputs to the outputs. The base algorithm set out in this paper is a deterministic function/mapping of the form:

$$(\boldsymbol{t}, \boldsymbol{u}^*, \boldsymbol{w}^*, n, r, T_*, T_\circ, T_\bullet) \mapsto (\boldsymbol{w}, \boldsymbol{u}, \boldsymbol{F}, \boldsymbol{U}).$$

Because it is a deterministic mapping, analysis using our queuing algorithm can be considered to be "distribution-agnostic" —i.e., so long as a model is within the G/G/$n$/G/+ model form we do not make distributional assumptions about quantities in the analysis. (Or to put it another way, to the extent that assumptions are made, they are assumptions of a deterministic nature.) Any stochastic aspects of the queuing analysis come in through an exogenous stochastic model for one or more of the inputs, and so the resulting stochastic properties of the queuing outputs are determined by simulation. If one were to extend the algorithm to allow for some additional stochastic elements (e.g., stochastic revival-times) we recommend using an additional input for any new stochastic element and ensuring that the internal mechanics of the algorithm remain deterministic. This ensures that the algorithm yields a deterministic mapping and will therefore retain its "distribution-agnostic" character.

## 5. Summary and conclusions

In this paper we have developed an algorithm for simulation of the G/G/$n$/G/+ model. Our algorithm converts any set of queuing inputs for this model into a full specification of outputs giving full information about the queuing process. Our algorithm is a deterministic function taking inputs for the arrival-times, intended use-times and patience-times for a set of users, plus a specification of the number of service facilities, their revival-time and their closure-times. Since the arrival-times, intended use-times and patience-times are inputs to the algorithm the method is distribution-agnostic — it can be used in conjunction with values for these inputs simulated from any joint distribution. Our algorithm leads to a full specification of the queuing process, and this can be summarised graphically in a "queuing plot" showing important time-metrics for each user and facility and a "queuing summary plot" showing the distributions of important time-metrics for the overall process. To assist in analysis of queuing problems, our algorithm and corresponding summary and plotting facilities are implemented in functions available in the `utilities` package in `R` (O'Neill 2021).

Our queuing algorithm is most useful in simulation analysis of queuing problems where we have a generative model for the inputs to the queuing process. In this context, we have shown that it is possible to simulate output metrics from the queuing process by taking in a set of simulations of the inputs. In particular, using simulated inputs from a generative model, it is simple to simulate the distributions of important output metrics such as the mean-waiting-times, mean-unserved-times and mean-use-times for the users. We have seen how this can be done over a range of values of $n$ specifying the number of service facilities in the amenity, which gives the analyst an understanding of how the distributions of important queuing metrics vary as we add or remove service facilities. This is useful in queuing analysis because it allows the analyst to examine the effect of varying the number of service facilities in an amenity. If the analyst is willing to stipulate costs for the service facilities, and costs for the total-waiting-time and total-unserved-time, it is then possible to generate the resulting distribution of losses, which allows computation of the loss distribution and the risk (expected loss) as a function of the number of service facilities. We have seen that this can be used to undertake optimisation of the number of service facilities.

The algorithm set out in this paper is valid for any $G/G/n/G/+$ model, which makes it generally applicable to a wide range of queuing problems. Nevertheless, this can be extended in various ways to allow more complicated user behaviour or specification of service facilities if that is desired. Some extensions would be quite simple and some would be more complicated. For any extended algorithm that yields a deterministic mapping, the distribution-agnostic character of the method will be retained. We hope that readers find our queuing algorithm and related plotting facilities to be useful for simulation analysis in queuing problems, and we welcome any extensions to this analysis.

**Supplementary materials**

All the graphical analysis and simulation analysis in this paper was undertaken in `R Studio` and is available in supplementary files. The available supplementary files are shown below:

```
Queuing Analysis.rmd              8KB
Queue Simulation Metrics.rds  1,577KB
```